\documentclass[11pt, oneside]{article}

\usepackage[a4paper, total={6.2in, 9in}]{geometry}
\pdfoutput=1
\usepackage{graphicx}
\usepackage{amssymb}
\usepackage{xcolor}
\usepackage[T1]{fontenc}
\usepackage{lmodern}
\usepackage[bottom]{footmisc}
\usepackage[normalem]{ulem} 

\PassOptionsToPackage{hyphens}{url}\usepackage[hidelinks]{hyperref}
\usepackage{fancyhdr}                           
\fancypagestyle{firstpage} 
{
   \fancyhf{}
   \fancyfoot[C]{\vspace{-12mm}\thepage}

   \fancyfoot[L]{\vspace{-5mm}\noindent\rule{6.2in}{0.4pt}
     \scriptsize{
         \textcopyright \hspace{0.5mm} Barclays Bank PLC 2016-2017 \\
       This work is licensed under a \href{https://creativecommons.org/licenses/by/4.0/}{Creative Commons Attribution 4.0 International License (CC
       BY).}  \\ Provided you adhere to the CC BY license, including as to attribution, you are
       free to copy and redistribute this work in any medium or format and remix, transform,
       and build upon the work for any purpose, even commercially.                  \\
       BARCLAYS is a registered trade mark of Barclays Bank PLC, all rights are reserved. \\
       \vspace{1mm}
       Electronic copy available at \href{http://arxiv.org/abs/1608.00771}{http://arxiv.org/abs/1608.00771}
     }
 }
}

\title{\vspace{-1cm}Smart Contract Templates: \\ foundations, design landscape and research directions}

\usepackage{anyfontsize}
\author{%
  \hspace{-1.5em}
  \begin{tabular}{c} {\fontsize{9.75}{1cm}\selectfont Christopher D. Clack} \\ {\fontsize{9.75}{1cm}\selectfont Centre for Blockchain
      Technologies} \\ {\fontsize{9.75}{1cm}\selectfont Department of Computer Science} \\ {\fontsize{9.75}{1cm}\selectfont University College
      London} \end{tabular} \hspace{-2.5em} \and
  \begin{tabular}{c} {\fontsize{9.75}{1cm}\selectfont Vikram A. Bakshi} \\ {\fontsize{9.75}{1cm}\selectfont Investment Bank CTO Office} \\
    {\fontsize{9.75}{1cm}\selectfont Barclays} \\ \hskip 1em \end{tabular} \hspace{-2em} \and
  \begin{tabular}{c} {\fontsize{9.75}{1cm}\selectfont Lee Braine} \\ {\fontsize{9.75}{1cm}\selectfont Investment Bank CTO Office} \\
    {\fontsize{9.75}{1cm}\selectfont Barclays}  \\ \hskip 1em \end{tabular} 
}

\date{August 4, 2016 \\ \footnotesize{(Revised March 15, 2017)}}

\begin{document}
\maketitle
\thispagestyle{firstpage} 
\vspace{-1cm}
\begin{abstract}
  In this position paper, we consider some foundational topics regarding smart contracts (such
  as terminology, automation, enforceability, and semantics) and define a smart contract as an
  automatable and enforceable agreement. We explore a simple semantic framework for smart
  contracts, covering both operational and non-operational aspects, and describe templates and
    agreements for legally-enforceable smart contracts, based on legal documents. Building upon
    the Ricardian Contract, we identify operational parameters in the legal documents
    and use these to connect legal agreements to standardised code.  We also explore the design
    landscape, including increasing sophistication of parameters, increasing use of common
    standardised code, and long-term research. 
\end{abstract}

\vspace{-5mm}
\section{Introduction}
\label{sec:introduction}

The aim of Smart Contract Templates \cite{smartcontracttemplates} is to support the management
of the complete lifecycle of ``smart'' legal contracts. This includes the creation of legal
document templates by standards bodies and the subsequent use of those templates in the
negotiation and agreement of contracts by counterparties. They also facilitate automated
performance of the contract and, in the event of dispute, provide a direct link to the relevant
legal documentation.

The templates and agreements may (or may not) be agnostic to the method by which a contract is
automated -- that is a design choice for the template issuer, counterparties, network, etc.
Smart legal contracts could potentially be implemented as software agents operating on a wide
range of technology platforms, including distributed ledger platforms such as \linebreak AxCore
\cite{axoni}, Corda \cite{corda2016}, Digital Asset Platform \cite{digitalasset}, Ethereum
\cite{ethereum}, and Fabric \cite{fabric}.

Here we aim to make a practical contribution of relevance to financial institutions.  We
consider how contracts are written, how they are enforced, and how to ensure that the automated
performance of a contract is faithful to the meaning of the legal documentation.  We discuss
these issues using reasonably straightforward language, so that it is accessible not only to
financial institutions but also to, for example, lawyers, regulators, standards bodies, and
policy makers.  We hope that the issues and views raised in this paper will stimulate debate
and we look forward to receiving feedback.

\vspace{3mm}
\noindent \textbf{Acknowledgements:} We would like to thank Clive Ansell (ISDA), Ian Grigg (R3)
and \linebreak Darren Jones (Barclays) for their helpful feedback.

\pagebreak
\section{Foundations}
\label{sec:background}
To lay the foundation for subsequent discussion, we elaborate four key topics of terminology,
  automation, enforceability, and semantics.

\subsection{Terminology --- ``smart contracts''}
In \cite{stark2016}, Stark gives an overview of the two different ways that the term ``smart
contract'' is commonly used:
\begin{enumerate}
\item The first is operational, involving software agents, typically but not necessarily on a
  shared ledger.  The word ``contract'' in this sense indicates that these software agents are
  fulfilling certain obligations and exercising certain rights, and may take control of certain
  assets within a shared ledger. There is no consensus on the definition of this use of the
  term ``smart contract'' --- each definition is different in subtle ways
  \cite{swansongreatchain, swanson2015, szabo1997}.  Stark renames these agents as {\em smart
    contract code}.
\item The second focuses on how legal contracts can be expressed and implemented in software.
  This therefore encompasses operational aspects, issues relating to how legal contracts are
  written and how the legal prose should be interpreted. There are several ideas and projects
  which focus on these aspects such as CommonAccord \cite{commonaccord}, Legalese
  \cite{legalese}, Monax's dual integration \cite{monax}, and the Ricardian
  Contract \cite{grigg2004ricardian}. Stark renames these as {\em smart legal contracts}.
\end{enumerate}

\noindent
Given that there is no clear consensus on the terminology being used, it is important that we
should be clear in this paper. We prefer that the term ``smart contract'' should cover both
versions, so we adopt a higher-level definition based on the two topics of automation and
enforceability (that are explored in depth in sections \ref{sec:automation} and
\ref{sec:enforceability}):\\

\begingroup
  \advance\leftmargini 2em
\begin{quote}

  {\bf
{\em A smart contract is an automatable and enforceable agreement.  Automatable by computer,
  although some parts may require human input and control.  Enforceable either by legal
  enforcement of rights and obligations or via tamper-proof execution of computer code.  }
  }
\end{quote}
\endgroup

\noindent
\\ This definition is sufficiently abstract to cover both ``smart legal contracts'' (where the
agreement is a legal agreement, at least some of which is capable of being implemented in
software) and ``smart contract code'' (which is automated software that may not necessarily be
linked to a formal legal agreement).  It simply states a requirement that the contract must be
enforceable without specifying {\em what} is the aspect being enforced; for smart legal
contracts these might be complex rights and obligations, whereas for smart contract code what
is being enforced may simply be the actions of the code.

We focus on smart legal contracts, with the expectation that they will be performed using smart
contract code.  Throughout the rest of this paper we also, for clarity, adopt Stark's terms
{\em smart contract code} and {\em smart legal contract}.

\pagebreak
\subsection{Automation}
\label{sec:automation}

We say that a smart contract is ``automatable'' rather than that it is ``automated'' because in
practice there may be parts of a legal agreement whose performance requires human input and
control.  However, to be a ``smart contract'' we require that some part of the agreement is
capable of being automated (otherwise it is not ``smart'').

Automation is generally accomplished by the use of one or more computers.  The phrase ``by
  electronic means'' is a synonym.  Our definition of smart contract does not require that this
  automation occurs on a shared ledger, though that is
  certainly a possible and even probable method.

As an example of how automation might be achieved using smart legal contracts, Grigg
\cite{grigg2015sumofchains} presents the Ricardian Contract triple of ``prose, parameters and
code''.\footnote{https://en.wikipedia.org/wiki/Ricardian\_Contract} The legal prose is linked
via parameters (name-value pairs) to the smart contract code that provides automation.  For
example, a software agent might have been developed that will be instantiated on a shared ledger
and, once initiated, will proceed to undertake various transfers of value in accordance with
the legal prose.  The parameters are a succinct way to inform the code of the final operational
details.

The code in this case would be suitable for a specific platform but we can imagine in the
future that multiple platforms could be targeted from a single contract.\footnote{This could be
  achieved by, for example, using the list of parameters to connect the legal prose to a {\em
    set} of smart software agents, e.g. one agent per platform.}

\subsection{Enforceability}
\label{sec:enforceability}

Given a smart contract must be ``enforceable'' \cite{nortonrose}, what are the
elements that must be enforced? And how?  First we consider {\em what} must be enforced:

\vspace{2mm}
\noindent
\subsubsection{What to enforce}

What needs to be enforced is different for smart contract code and smart legal contracts:

\begin{itemize}

\item For \textbf{{\em smart contract code}}, the key requirement is that the code should run
  successfully and accurately within a reasonable time.  If the technology platform is in
  complete control of all of the actions of the smart contract code then these actions should
  occur faithfully and without undue delay. Things that could go wrong (and therefore require
  ``enforcement'') include technical issues within the platform and issues that take
  place outside of the platform --- an obvious example would be the physical delivery of goods.

\item For \textbf{{\em smart legal contracts}}, things can be considerably more complex.
  Typically a legal contract would include rights and obligations that accrue to the different
  parties and are legally enforceable. These are often expressed in complex, context-sensitive,
  legal prose and may cover not just individual actions but also time-dependent and
  sequence-dependent sets of actions.  There may also be overriding obligations on one or more
  of the parties such that a {\em lack} of action could be deemed to be a wrong-performance or
  non-performance of the contract.
\end{itemize}

\subsubsection{How to enforce}

Enforcement might be achieved via traditional or non-traditional methods:

\begin{itemize}
\vspace{-0.5mm}
\item \textbf{{\em Traditional}} means of enforcement include a variety of dispute resolution
  methods such as binding (or non-binding) arbitration, or recourse to the courts of law.
  There is an established body of law, and the methods by which parties can resolve disputes
  are well known.  For illegal acts, courts are for example empowered (to different extents,
  according to jurisdiction) to impose fines, sequester assets, or deprive the wrong-doer of
  liberty.  For disputes relating to contracts, the courts have extensive experience of
  adjudicating on issues of contract wrong-performance and non-performance, of
  awarding damages or other reliefs as appropriate, and in some cases assisting in the
  enforcement of payment of damages.

  \item \textbf{{\em Non-traditional}} methods of enforcement may also be imagined.  For
    example, there is currently debate and experimentation on enforcing the actions of smart
    contract code at a network level without the need for dispute resolution. This is a
    fundamentally different notion of enforcement that is often expressed in terms of
    ``tamper-proof'' technology, with the assumption that in a perfect implementation of the
    system wrong-performance or non-performance become impossible.
  \item[] ``Tamper-proof'' technology is typically described in terms of distributed networks
    of computers that are unstoppable and in a technological sense cannot fail regardless of
    malicious acts, power cuts, network disruption, natural catastrophies or any other
    conceivable event. For example, a ``permissionless'' shared ledger might make use of
    tamper-proof technology. Swanson \cite{swanson2015} gives a good overview of many of the
    complex issues that arise with permissioned and permissionless distributed consensus
    systems. With such a system, it is assumed that a software agent, once started, could not
    be stopped. For truly ``unstoppable'' software agents, code must be defined to take the
    appropriate action in response to various dynamic states that might occur (such as another
    party defaulting on a required payment).  In a truly unstoppable ``tamper-proof'' version
    of the system, all such possibilities would have to be anticipated and appropriate actions
    determined in advance.
\end{itemize}

\noindent
Although some groups are actively pursuing tamper-proof smart contract code, our preference is
for smart legal contracts that are enforceable by traditional legal methods for reasons
including:
\begin{itemize}
\item In a system with enforcement by tamper-proof network consensus, there would be no
  ``executive override'' provisions.  Agreements, once launched as smart contract code, could
  not be varied.  But it is common for provisions of an agreement to be varied dynamically ---
  for example, to permit a client to defer paying interest, or to permit a payment holiday, or
  to permit the rolling-up of interest over a period.  Unless every possible variation is coded
  in advance, none of this would be possible in a tamper-proof system.
\item Enforcement by network consensus can only apply to the fulfilment of obligations, or the
  exercising of rights, that are under the control of the network. However, objects and actions
  in the physical world are unlikely to be under full (if any) control of the network.
\item Mainelli and Milne \cite{mainelli2016} observe that smart contract code ``that involved
  payments would require posting collateral to be completely automated. This locking-up of
  collateral would lead to a serious reduction in leverage and pull liquidity out of
  markets. Markets might become more stable, but the significant reduction in leverage and
  consequent market decline would be strongly resisted by market participants.''
\end{itemize}

\subsection{The semantics of contracts}
Part of our remit is to consider the semantics of a contract --- i.e. what is the ``meaning''
of a contract?  We view a legal contract as having two aspects:
 
\begin{enumerate}
\item The {\bf \em operational aspects:} these are the parts of the contract that we wish to
  automate, which typically derive from consideration of precise actions to be taken by the
  parties and therefore are concerned with performing the contract.

\item The {\bf \em non-operational aspects}: these are the parts of the contract that we do not
  wish to (or cannot) automate.
\end{enumerate}
 
\noindent
We may approach the semantics of these two aspects of the contract in different ways. For
example, with the operational aspects we may wish to compare a semantic analysis of the
contract with a semantic analysis of the computer code --- if it were possible to develop a
proof for semantic equivalence\footnote{Approaches to formal semantics include, for example,
  denotational semantics and operational semantics.}, this could be used early in the
development lifecycle to increase confidence and reduce testing and debugging effort.  By
contrast, for the non-operational aspects of the contract we may wish to conduct a range of
different semantic analyses --- e.g. to analyse different forms of risk associated with a
contract.
 
A contract may comprise several documents, and the process by which these documents are agreed
may be complex. The semantics of the non-operational aspects of even quite straightforward
contracts can be very large and complex, yet by contrast the semantics of the operational
aspects might be simple and easily encoded for automation.
 
The operational aspects of a contract would typically dictate the successful performance of the
contract to completion.  If a dispute arises, then the non-operational aspects of the contract
would typically dictate what happens next --- i.e. in the context of the rights and obligations
of the parties, the specification of what remedies shall be applied in the case of
contract partial-performance or non-performance by one party.
 
The greater part of a legal contract may often be devoted to defining the rights and
obligations of the parties in the event of a problem.  Sometimes, the actions to be taken in
case of material breach of contract are expressed precisely; however, this is not always the
case and dispute resolution may require a protracted process of negotiated settlement,
arbitration, or court proceedings.
 
Furthermore, it is important to understand the role of law. A lawyer would read and understand
the contract {\em in the context of} the governing law --- i.e. each legal document must be
interpreted according to the relevant law (corporate law, consumer law, etc.) of its stated or
inferred jurisdiction, and therefore the semantics of that law must also be understood.  It
should be noted that the issue of law relates not only to the non-operational aspects but also
to the operational aspects --- for example, trading with certain countries may be illegal due
to government-imposed sanctions.

\pagebreak
Given this semantic framework for the legal contracts that underpin financial instruments, we
can derive a different perspective of smart contracts:
 
\begin{itemize}
\item smart contract code focuses exclusively on automation by computer and therefore concerns
  itself only with performing those {\em operational aspects} that are expressed in the code,
    whereas
\item smart legal contracts consider both the operational and non-operational aspects of a
  legal contract, some of whose operational aspects must then be automated (presumably by smart
  contract code).
\end{itemize}
 
\begin{figure}[h]
\begin{center}
\includegraphics[height=7cm]{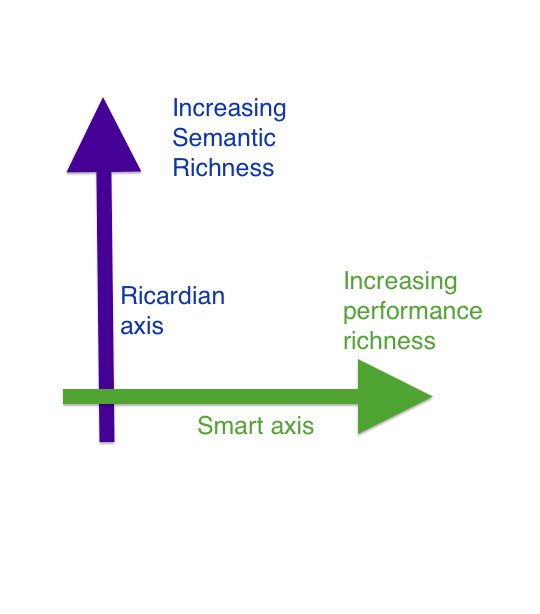}
\parbox{4.5in}{\vspace{-3cm}\caption{\footnotesize From Grigg \cite{griggintersection}, the y-axis
    represents an increasing ability to capture the semantics of a smart legal contract,
    whereas the x-axis represents the increasing performance of smart contract code.
\label{fig:2D}}}
\end{center}
\end{figure}

\noindent
This idea was previously expressed in a slightly different way by Grigg
\cite{griggintersection}, displayed as a chart where the y-axis was denoted the ``Ricardian
axis'' of increasing semantic richness (i.e. increasingly capturing the semantics of the legal
prose) and the x-axis was the ``Smart axis'' of increasing performance richness (primarily
concerned with the execution of smart contract code): see Figure~\ref{fig:2D}. Both are
  important, yet they are orthogonal issues and with appropriate interfacing developments can
  potentially proceed along both axes simultaneously.

\section{Smart Contract Templates}
Smart Contract Templates provide a framework to support complex
legal agreements for financial instruments, based on standardised templates. Following Grigg's
Ricardian Contract triple \cite{grigg2015sumofchains}, they use parameters to connect legal
prose to the corresponding computer code, with the aim of providing a legally-enforceable
foundation for smart legal contracts.

Complex sets of legal documentation can be augmented with the identification of operational
parameters that are key to directing the behaviour of the smart contract code (in this paper we
call these ``code parameters'') --- the smart contract code is assumed to be {\em standardised}
code whose behaviour can be controlled by the input of such parameters.

Here we explore the design landscape for the implementation of Smart Contract Templates.  We
observe that the landscape is broad and that there are many potentially viable sets of design
decisions.

\subsection{Templates and Parameters}
\begin{figure}[h]
\begin{center}
  \includegraphics[height=5cm]{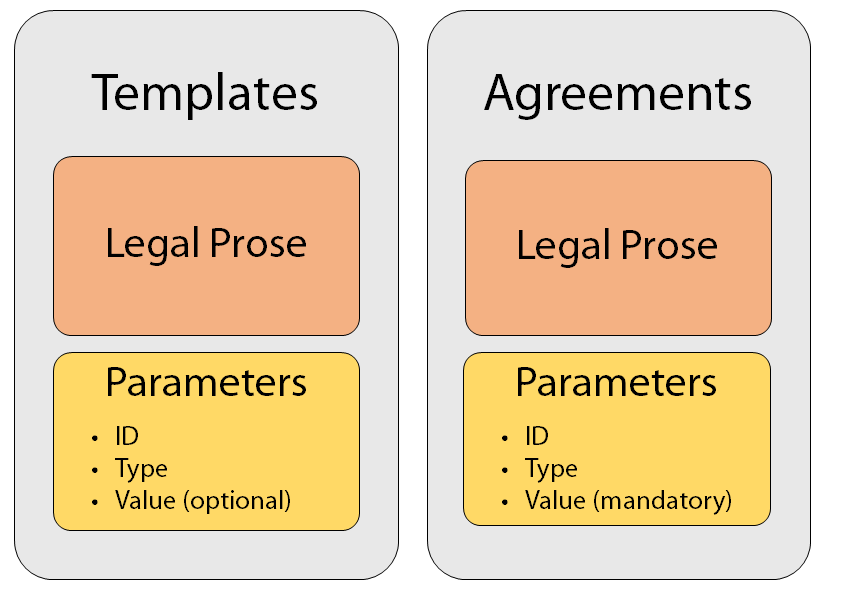}
  \parbox{4.5in}{\break\caption{\footnotesize {A template may contain both legal prose and
        parameters. Each parameter has an identifier (a name), a type, and an optional
        value. Agreements are derived from templates, and both the legal prose and parameters
        may be customised during negotiation. Values are mandatory for all parameters in a
        signed agreement.
\label{fig:SCT-1}
}}}
\end{center}
\end{figure}

A template is an electronic representation of a legal document as issued by a standards body
--- for example, by the International Swaps and Derivatives Association (ISDA). As illustrated
in Figure~\ref{fig:SCT-1}, a template contains both legal prose and parameters, where each
parameter has an identity (a unique name), a type, and may (but need not) have a value.

\vspace{4mm}

\noindent
An agreement is a fully-instantiated template (including any customised legal prose and
parameters).  The customisation of legal prose and parameters at this stage is commonplace and
results from negotiation between the counterparties. The legal prose of an agreement will be
derived from that of the template, but need not be identical, and similarly the parameters of
the agreement will be derived from the template but need not be identical.  It is also common
for agreements to comprise multiple documents such as framework agreements (e.g. a Master
Agreement) with various annexes (e.g. a Schedule) and credit support documentation (e.g. a
Credit Support Annex).

\vspace{1mm}
Deriving the set of code parameters may be complicated by three factors:
\begin{enumerate}
\item It is common for parameters to be embedded in the legal prose --- such parameters would
  initially be identified visually, aided by a graphical user interface.
\item Some of the values identified as ``parameters'' in the agreement (or template) may not
  have an operational impact and therefore would not be passed to the smart contract code.
\item It is possible for a parameter to be defined (given a name) in one document, given a
  value in a second document, and used (e.g. in business logic) in a third document.
\end{enumerate}
Although parameters need not have values in a template, they must have values in a signed
agreement. All of an agreement's parameter values are a critical part of the contract as they
directly reflect the business relationship between parties and those that are code parameters
influence the automated operation of the contract.

\subsection{The design landscape for Smart Contract Templates}
In this section, we consider the possible areas for future development of Smart Contract
Templates. We do this by considering three areas of development relating to: \linebreak (i) the
sophistication of parameters, (ii) the standardisation of code, and (iii) long-term research.

\subsubsection{Increasing the sophistication of parameters}
\begin{figure}[h]
\begin{center}
\includegraphics[height=3cm]{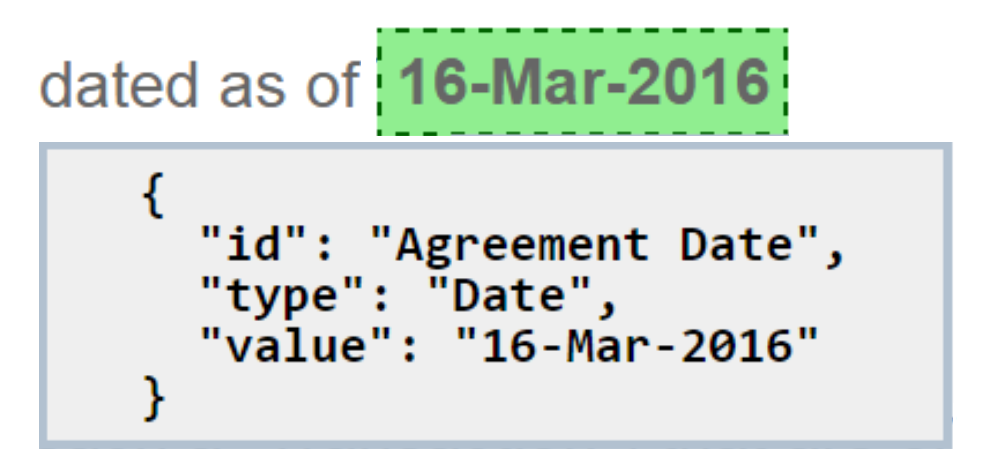}
\parbox{4.5in}{\break\caption{\footnotesize {From a Barclays demonstration of Smart Contract
     Templates: \protect\linebreak an editor permits a date in the legal prose to be highlighted, and then
     annotated to denote a simple parameter.  The parameter has a name ``Agreement Date'',
     type ``Date'' and value ``16-Mar-2016''.
\label{fig:SCT-POC-basetype}}}}
\end{center}
\end{figure}

Most parameters in existing templates have simple types, such as date, number, etc. These are
``base'' or ``primitive'' types\footnote{\url{https://en.wikipedia.org/wiki/Data_type}} and, as
an example, Figure~\ref{fig:SCT-POC-basetype} illustrates the identification
of a date in a master agreement; once highlighted and annotated, the name (``Agreement Date''),
type (``Date'') and value (``16-Mar-2016'') of this parameter will be passed to the code.

\vspace{3mm}

\vspace{-2mm}
\noindent

It is not necessary for parameters to be restricted to base types.  It is very likely that
values of more complex types, such as lists, will also need to be transferred to the code.

The passing of parameters to the code is necessary because of the desire to use {\em
  standardised} code. The number of parameters, and the complexity of the types of those
parameters, will typically increase as the code becomes more generic.

Beyond parameters with base types and more complex types such as lists, parameters can also be
expressions containing references to other parameter names. Unless those other parameter names
are defined within the expression, the expression is effectively a function.  Where a function
is passed as a parameter, this is known as a ``higher-order'' parameter and the receiving code
is known as a ``higher-order''
function.\footnote{\url{https://en.wikipedia.org/wiki/Higher-order_function}}

\begin{figure}[h]
\begin{center}
\includegraphics[height=1.75cm, width=10cm]{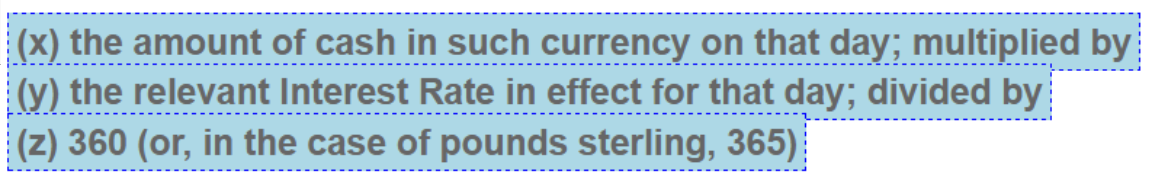}
\break
\includegraphics[height=1.5cm]{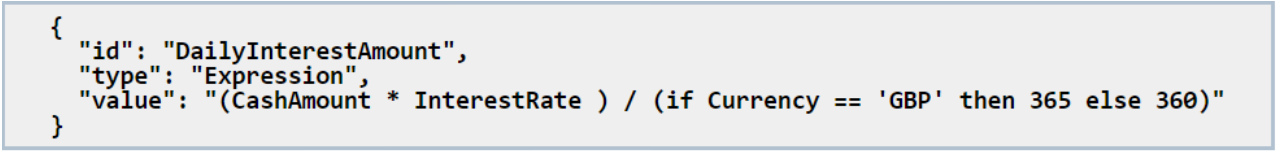}

\parbox{4.5in}{\break\caption{\footnotesize {From a Barclays demonstration of Smart Contract
      Templates: an editor permits business logic from the legal prose to be highlighted, and
      then annotated to denote a higher-order parameter. The parameter has a name
      ``DailyInterestAmount'', type ``Expression'', and value corresponding to an arithmetical
      expression.
\label{fig:SCT-POC-expressiontype}}}}
\end{center}
\end{figure}

An example of a higher-order parameter is illustrated in
Figure~\ref{fig:SCT-POC-expressiontype} where some business logic in the legal prose has been
highlighted and annotated to be a parameter with name ``DailyInterestAmount'' of type
``Expression''\footnote{In a conventional type system this expression would have an associated
  function type such as: \linebreak (Decimal, Decimal, Currency) -$>$ Decimal} and with a value
that is an encoding of the business logic in a format that is easily understandable by a
computer.  This business logic refers to three things whose values are unknown. The first two
are simple: ``the amount of cash in such currency on that day'' (in the expression this is
called CashAmount), and ``the relevant Interest Rate in effect for that day'' (in the
expression this is called InterestRate). The third occurs in the phrase ``in the case of pounds
sterling'', which requires some analysis to determine that it is referring to the prevailing
currency (hence the name Currency used in the expression) and that the normal abbreviation for
pounds sterling is ``GBP''.  Since this expression contains three parameter names whose values
are unknown, it will be stored in the set of code parameters as a function taking three
arguments (CashAmount, InterestRate, and Currency) and returning a numeric value that is the
result of the expression.

When business logic is converted into an expression this may involve the creation of new
parameter names (e.g. a new name for the expression itself, and for unknown quantities).
Sometimes the business logic may refer to a name that is already defined as a parameter, and
sometimes it may refer to a value provided by an ``oracle'' --- i.e. a value, such as an
interest rate, that is provided from a trusted source of data and is available to the code
while it is running.

The use of parameters may not only support greater standardisation of code.  In the far future,
we may see an increasing use of a formally structured style of expression embedded in legal
prose; if all business logic in legal prose could be replaced with arithmetical or logical
expressions, such as the higher-order parameters discussed in the previous paragraph, this
should lead to reduced ambiguity in legal prose and fewer errors. The adoption of formal logic
into legal prose would require such formal constructs to gain acceptance in the courts and to
be admissible as evidence of the intentions of the parties.

\begin{figure}[h]
\begin{center}
\includegraphics[height=2.5cm, width=15cm]{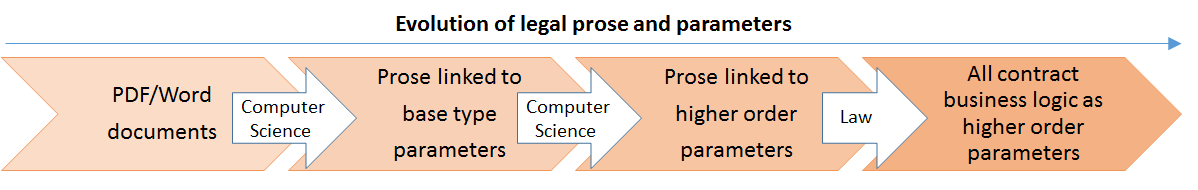}
\parbox{4.5in}{\break\caption{\footnotesize{ Parameters may become more sophisticated in the
      future, evolving from just simple base type parameters to also include more complex
      higher-order parameters. In the far future, if the encoding of business logic used in the
      parameters becomes acceptable to lawyers and admissible in court, then it could
      potentially replace the corresponding legal prose.
     \label{fig:progression-parameters}}}}
\end{center}
\end{figure}

Figure~\ref{fig:progression-parameters} illustrates how the sophistication of parameters and
their role in Smart Contract Templates may evolve in the future.

\vspace{3mm}

\vspace{-3mm}
\subsubsection{Increasing the use of common standardised code}
\label{sec:code-standardization}

In the previous subsection, we observed that the passing of parameters to smart contract code
is necessary because of the desire to use {\em standardised} code. This is important for
efficiency reasons as different smart contract code would otherwise have to be built, tested,
certified and deployed for {\em every} different trade. The effort is reduced if such code can
be standardised with parameters being passed to each invocation of that code.

This drives a desire for greater genericity of code, which can be enabled by passing more
parameters, and/or more sophisticated parameters (with more complex types).  Yet there remains
the problem that each bank currently manages its own distinct codebases. If smart contract code
could be {\em common} (i.e. {\em shared}) then it could be built, tested and certified once ---
and then utilised by every counterparty.

\begin{figure}[h]
\begin{center}
\includegraphics[height=2.5cm, width=15cm]{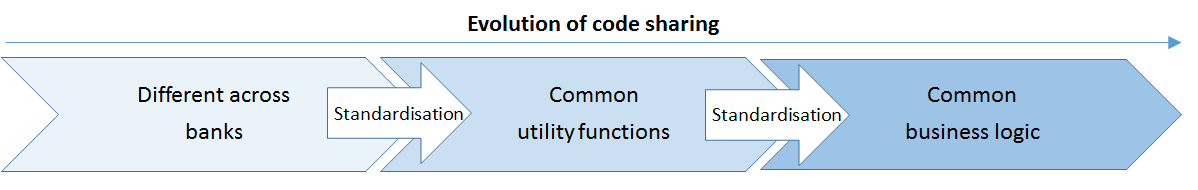}
\parbox{4.5in}{\break\caption{\footnotesize {Code may become more standardised in the future
     through increased sharing, evolving from different codebases across banks to greater
     adoption of common utility functions to common business logic.
\label{fig:progression-code}}}}
\end{center}
\end{figure}

We envisage that the potential economic benefits of using common (shared) code will drive
greater adoption in the future. One possible evolutionary route could build upon the use of
common utility functions --- programs that are already very nearly identical in all
counterparties. As the potential economic benefits become clearer and the supporting
technologies mature, the size and importance of such common code could increase until,
eventually, common business logic may become standardised smart contract
code. Figure~\ref{fig:progression-code} illustrates how the sharing of code may evolve in the
future.

\vspace{3mm}

\subsubsection{Long-term research challenges}

Several research challenges concerning smart contracts, shared ledgers, and blockchains are
currently being explored in academia. Financial institutions are providing input and
inspiration by highlighting relevant business challenges in technology and operations. A good
example is the potential to increase straight-through processing. Currently, lawyers draft
legal contracts, which are then negotiated and changed by possibly other teams of lawyers, and
then operations staff inspect the contract documents and/or other materials\footnote{Other
  supporting materials may include confirmations, emails, facsimiles, telephone recordings
  etc.} to identify the parameters that are then passed to code that may have been written some
time ago.

This raises several issues:
\begin{itemize}
\item Can we be absolutely certain of the meaning of the contract?  Are all parties truly
  agreed on the meaning of the contract, or do they instead each have a subtly different
  understanding of what the contract means?

\item Can we be certain that {\em all} code parameters have been identified, and that each is
  operationally relevant?  And can we be certain that their names, types and values have been
  faithfully transcribed?

\item After the parameters have been passed to the code, and the code runs, can we be certain
  that the code will be faithful to the semantics of the operational aspects of the contract?
  And will it do so under all conditions?

\end{itemize}

\noindent
A possible solution would be to develop a formal language in which to write legal documents ---
i.e. contract documents --- such that the semantics would be clear and the code parameters
could automatically be identified and passed to standardised code (alternatively, new code
could be generated).  This formal language would:
\begin{enumerate}
\item derive a number of important qualities from well-designed computer programming languages,
  such as a lack of ambiguity, and a compositional approach where the meaning of any clause can
  be clearly deduced without reading the rest of the document; and
\item be simple and natural to use, to such an extent that a lawyer could draft contracts using
  this formalism instead of using traditional legal language.
\end{enumerate}

\noindent
The former aspect has already received considerable attention in academia (see survey in
\cite{hvitved12phd}) and beyond (e.g. the open-source Legalese project).  In contrast, the
latter aspect is likely to be by far the greater challenge.

Another challenge is whether such a contract, written in a computer-like language, would be
admissible in court as a true and faithful representation of the intentions of the parties.
Issues of signature and tamper-evident documents are easily solved, yet whether a court would
accept the definitions of the meanings of phrases in such a contract is not immediately clear.

\begin{figure}[h]
\begin{center}
\includegraphics[width=15cm]{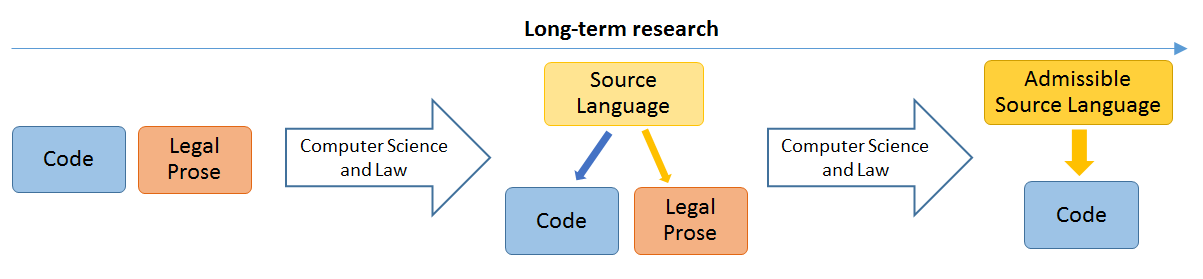}
\parbox{4.5in}{\break\caption{\footnotesize {Long-term research may lead from existing separate
      code and legal prose to source languages which can be automatically translated into both
      code and legal prose, with the prose being admissible in court. Even longer term research
      could result in formal languages which themselves are admissible in court. Note, this
      figure omits parameters for clarity.
\label{fig:progression-research}}}}
\end{center}
\end{figure}

As illustrated in Figure~\ref{fig:progression-research}, this problem could be
solved in two ways:
\begin{enumerate}
\item Initially, the language could generate a document version of the contract in a more
  ``natural'' legal style, with the expectation that this document would be admissible in
  court.
\item Eventually, further research in domain-specific languages and law could result in a new
  formalism itself being admissible in court.
\end{enumerate}

\subsubsection{Future developments and initial requirements}

\begin{figure}[h]
\begin{center}
\includegraphics[width=14cm]{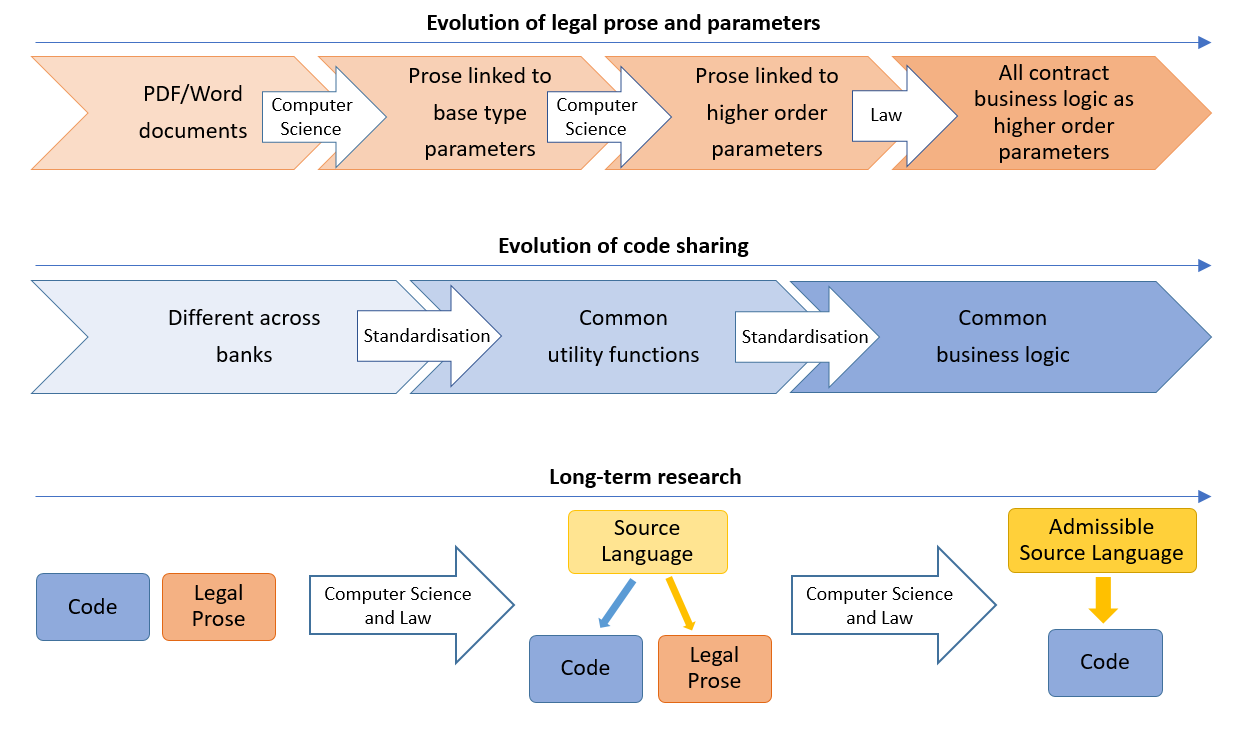}
\parbox{4.5in}{\break\caption{\footnotesize {Potential evolution of aspects of
      legally-enforceable smart contracts in three streams: legal prose and parameters, code
      sharing and long-term research.
      \label{fig:roadmap}}}}
\end{center}
\end{figure}

\noindent The areas of future development described in the preceding sections are brought
together in Figure~\ref{fig:roadmap}, illustrating the potential evolution of aspects of
legally-enforceable smart contracts.

\vspace{-5mm}
\noindent
This complexity motivated us to sketch the following initial set of requirements for the design
landscape of storage, processing and transmission of smart legal contracts:

\begin{itemize}
\item support both legal prose and parameters;
\item support structured data formats such as XML, FpML, etc.;
\item support the instantiation of multiple agreements from a single template;
\item permit parameters to be defined in one document, given a value in a second document, and
  used in a third document;
\item support contracts that comprise multiple documents;
\item support a wide range of parameter types, including higher-order parameters in future;
\item support the extraction of code parameters for transmission to the smart contract code;
\item support increasing standardisation and sharing of common code;
\item support increasing automation of, and interaction between, legal prose and code;  
\item support multiple execution platforms;
\item support Ricardian Contracts (and therefore, for example, support digital signing of
  contracts and the construction of a cryptographic hash of the contract).
\end{itemize}

\noindent
 We aim to report on progress in a subsequent paper, including essential requirements and
 design options of potential formats for the storage and transmission of smart legal
 contracts.

\vspace{5mm}
\section{Summary and Further Work}

\subsection{Summary}
This paper has considered four foundational topics regarding smart contracts: terminology,
automation, enforceability, and semantics. We defined a smart contract as an automatable and
enforceable agreement. We viewed legal contracts within a simple semantic framework based on
two key aspects: one being the operational aspects concerning automation of the contract and
another being the non-operational aspects.  We then described templates for legally-enforceable
smart contracts as electronic representations of legal documents containing prose and
parameters (with each parameter comprising an identifier, a type and an optional
value). Agreements are then fully-instantiated templates, including customised legal prose and
parameters. By selecting the appropriate smart contract code, this
approach results in the creation of Ricardian Contract triples.

We then explored the design landscape: increasing the sophistication of parameters to complex
higher-order types and business logic that could be admissible in court and potentially replace
the corresponding legal prose. We also explored increasing the use of common standardised code
through greater sharing, evolving from different codebases across banks to broader adoption of
common utility functions to common business logic. Additionally, long-term research was
outlined which could lead to source languages which can be automatically translated into both
code and legal prose; even longer term research could result in formal languages which
themselves are admissible in court.

\subsection{Further Work}

\noindent
A benefit of looking to the future is that it helps to identify a potential roadmap for
applying research within industry. Smart Contract Templates have already demonstrated a way to
link {\em standardised agreements} to {\em standardised code} and so, in the near term, it may
be possible to utilise them with existing infrastructure. In the longer term, they could be
utilised on shared ledgers.

We are continuing to collaborate broadly, including via trade associations such as ISDA and
FIA, on exploring formats for the storage and transmission of smart legal agreements comprising
legal prose and parameters. Of particular interest are the corresponding essential requirements
and design options, including the possibility of extending the scope of existing data standards
such as FpML (Financial products Markup Language) for derivatives.

There are many open questions for the future.  We have explored some of these questions in this
paper, but we will end with one more: is it possible to provide straight-through-processing of
financial contracts, with full confidence in the fidelity of the automated computer code to the
entire semantics of the contract? This, of course, will require substantial work and
collaboration by the financial services industry, standards bodies, academia\footnote{ For
  example, further investigation of contract semantics and the specification language CLACK is
  being pursued at University College London.}, and lawyers.

\pretolerance=-1
\tolerance=-1
\emergencystretch=0pt

\pagebreak

\bibliography{SCT2016}
\bibliographystyle{plain}

\end{document}